\documentclass[
 reprint,
 amsmath,amssymb,
 aps, prl,
]{revtex4-2}

\usepackage{morefloats}
\usepackage{color}
\usepackage{epsfig,graphicx,amsfonts,amsbsy}
\usepackage{amsmath,amsfonts,amsthm,amssymb}
\usepackage{appendix}
\usepackage{bbm}
\usepackage{makeidx}
\usepackage{url}
\usepackage{verbatim}
\usepackage{mathrsfs} 
\usepackage{morefloats}
\usepackage{appendix}
\usepackage{bbm}
\usepackage{makeidx}
\usepackage{url}
\usepackage{verbatim}
\usepackage[bookmarksnumbered,pdfpagelabels=true,plainpages=false,colorlinks=true,linkcolor=blue,citecolor=blue,urlcolor=blue]{hyperref}
\usepackage[bookmarksnumbered,pdfpagelabels=true,plainpages=false,colorlinks=true,linkcolor=blue,citecolor=blue,urlcolor=blue]{hyperref}
\usepackage{array}
\usepackage{booktabs}
\usepackage{multirow}
\usepackage{bbm}
\usepackage{tabularx}
\usepackage{cancel,soul}
\usepackage{nicefrac, xfrac}
\usepackage{ulem}
\usepackage{bbold}

\newcommand{\sectioncustom}[1]{\noindent {\it #1.- }}

\usepackage{etoolbox} 

\def \fcfm {Departamento de F{\'i}sica, Facultad de Ciencias Físicas y Matemáticas, Universidad de Chile, Santiago, Chile.}

\begin{document}

\title{  Antiferromagnetic Hall-Memristors}
\author{Gaspar De la Barrera} 
    \email{gaspar.delabarrera@ug.uchile.cl}
\author{Alvaro S. Nunez}
\email{alnunez@dfi.uchile.cl}
\affiliation{$^1$\fcfm}
\date{\today}

\begin{abstract}
 Spin-memristors are a class of materials that can store memories through the control of spins, potentially leading to novel technologies that address the constraints of standard silicon electronics, thereby facilitating the advancement of more intelligent and energy-efficient computing systems. In this work, we present a spin-memristor based on antiferromagnetic materials that exhibit Hall-memresistance. Moreover, the nonlinear Edelstein effect acts as both a writer and eraser of memory registers. We provide a generic symmetry-based analysis that supports the viability of the effect. To achieve a concrete realization of these ideas, we focus on CuMnAs, which has been shown to have a controllable nonlinear Hall effect. Our results extend the two-terminal spin-memristor setting, which is customarily the standard type of device in this context, to a four-terminal device.
\end{abstract}

\maketitle

\sectioncustom{Introduction} Memristors were conceived by Leon Chua in 1971\cite{Chua1971}. Their defining characteristic is "memristance," a resistance that depends on the history of the current or voltage applied to the device\cite{Strukov2008}. This unique property allows memristors to "remember" the amount of charge flowing through them, making them promising candidates for nonvolatile memory\cite{Tetzlaff2013, Chua2019}, neuromorphic computing\cite{Kim2025, Duan2024, Adamatzky2013}, and analog circuit applications\cite{Yang2012}. Fabricated using various materials and structures, the practical realization of solid-state memristors by Hewlett-Packard in 2008 sparked significant interest in their potential to overcome limitations in traditional computing architectures\cite{Waser2012}.

Spintronic memristors, which utilize magnetic tunnel junctions, nanomagnet ensembles, domain walls, topological spin textures, and spin waves, have been introduced elsewhere\cite{Shao2025, Lequeux2016, Mansueto2021}. Each embodies distinct state spaces. These memristors can display steady, oscillatory, stochastic, and chaotic behavior within their state spaces, enabling applications for in-memory logic, neuromorphic computing, and stochastic and chaos-based computing.
A spin-memristor entails a novel fusion of spintronics with memristive technology\cite{Jaro2023, Qin2023}, blazing the trail for advanced computing and memory systems. Unlike traditional memristors that depend solely on electrical charge flow, spin memristors utilize electron spin characteristics to adjust resistance and maintain information. This combined use of charge and spin properties boosts energy efficiency, scalability, and data retention, positioning spin memristors as a strong contender for future neuromorphic computing, nonvolatile memory, and spintronic logic technologies.
The resistance of a spintronic device is affected by its magnetization state. 
The overall magnetization distribution determines their resistance.

A parallel line of research has led to the development of antiferromagnetic (AFM) spintronics\cite{MacDonald2011}, a rapidly growing field that harnesses the unique properties of antiferromagnetic materials for next-generation spintronic devices. Unlike ferromagnets, which feature aligned electron spins, antiferromagnets exhibit alternating spin directions, resulting in no net macroscopic magnetization. This crucial difference offers several advantages, including ultrafast dynamics in the terahertz range, which are orders of magnitude faster than those of ferromagnets \cite{Jungwirth2016, DalDin2024, Fukami2020, Baltz2018}. The absence of stray magnetic fields enables higher integration density without interference between neighboring bits, resulting in more compact and stable memory and logic architectures. Their robustness against external magnetic fields ensures reliable operation in various environments, and their thermal stability at higher temperatures, due to high Néel temperatures, is beneficial for industrial applications. Furthermore, AFMs show potential for lower energy consumption in switching and maintaining spin states, thereby contributing to the development of sustainable electronics. AFMs also exhibit unique transport phenomena, such as anisotropic magnetoresistance, the spin Hall effect, and the anomalous Hall effect, which can be used to read out and manipulate their spin states electrically.

This work proposes a memristor constructed from antiferromagnetic components that exhibit Hall-memristance.
In terms of the mathematical description of this {\bf antiferromagnetic memristor}, our ideas are expressed as:
$$
J^{y}=\sigma(w) E^2_x\;\;\;\;\;\;\;\;\;\;\;\; \frac{dw}{dt}=g(w, J^{y})
$$
where $J^{i}$ is the current density, $\sigma(w)$ is the non-linear-Hall-memconductance\cite{Du2021} that depends explicitly on an internal parameter, $w$, and  $ E_k$ is the electric field.  Here, $g$ is a generic function that describes the evolution of the internal parameter coupled to the current.
To achieve a concrete realization of these ideas, we will explore a variety of antiferromagnetic materials; the most promising at this point is  CuMnAs, which has been shown to have a controllable nonlinear Hall effect in \cite{Hu2024}. Concerning $w$, we focus on intrinsic antiferromagnetic degrees of freedom, such as Néel field order parameter strength. These are altered through the nonlinear Edelstein effect. The physical origin of the coupling $g$ is the aforementioned magnetoelectric coupling standard in antiferromagnetic materials.

\begin{figure}
    \centering
    \includegraphics[width=1\linewidth]{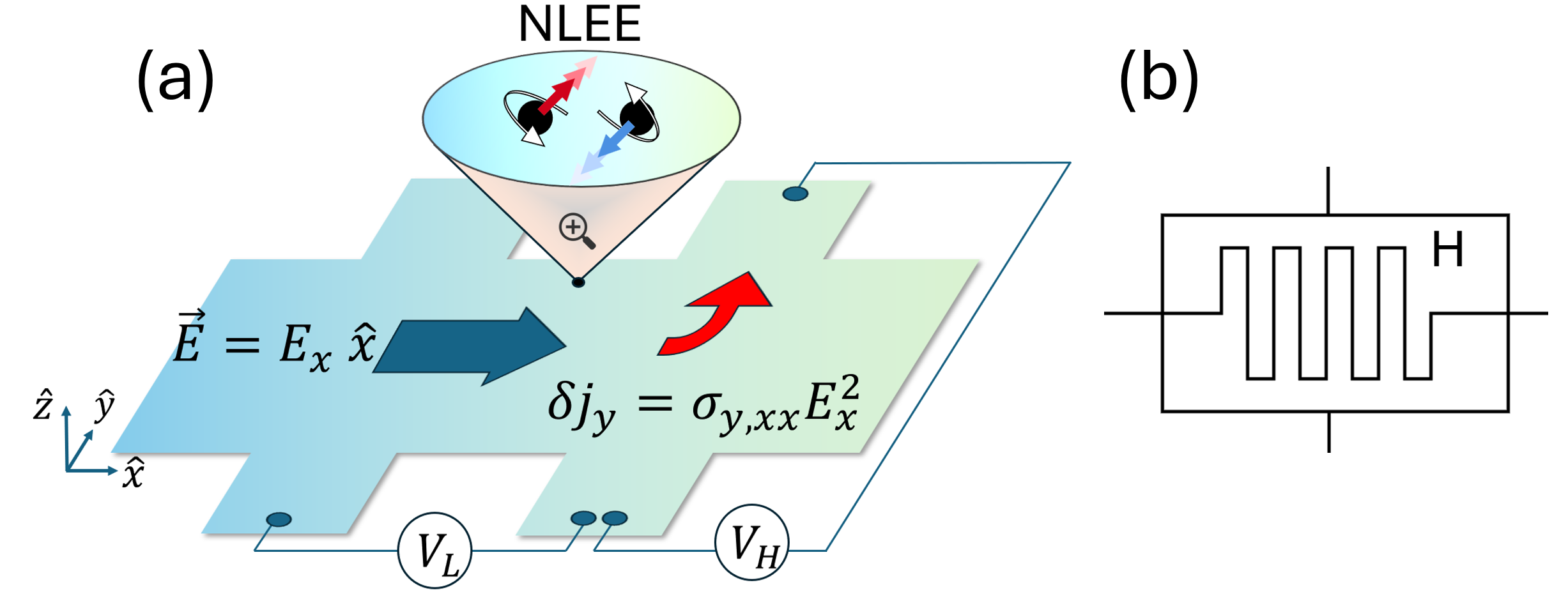}
    \caption{Schematic picture of the antiferromagnetic Hall-memristor. An electric field is applied in the $x$-direction through a voltage $V_L$, then a transversal current is generated and measured through a transversal voltage $V_H$. The memory effect is caused by the NLEE, which alters the Néel vector in the material when an electric field is applied, thereby changing the magnetic coupling and transport properties of the material. (b) shows our proposed circuit diagram symbol for the Hall-memristor.}
    \label{fig:schematic}
\end{figure}

\sectioncustom{Nonlinear Hall and Edelstein effects} Key concepts in the following paragraphs are encoded in the physical nature of the nonlinear Hall and Edelstein effects.
The Hall effect is concerned with a transverse voltage that appears when a current flows through a material in a direction perpendicular to the direction of the current. These effects are significant in condensed matter physics, leading to major discoveries, such as topological phases of matter and fractional charges. Typically, the Hall effect occurs when time-reversal symmetry is broken, often by the presence of magnetic fields. To get a Hall effect without breaking this symmetry, we can either look at spin or valley properties, or go beyond the usual linear response. The recently discovered nonlinear Hall effect (NLHE) is an example of the latter\cite{Du2021, Du2021a, Huang2023, Yamaguchi2024}. Unlike alternative Hall effects, where voltage and current have a simple linear relationship, the NLHE involves a transverse voltage that depends nonlinearly (specifically, quadratically) on the longitudinal current. This effect works by breaking inversion symmetry, rather than time-reversal symmetry. This opens up exciting new ways to study quantum materials and can even be applied to other unusual responses when different symmetries are broken, including the nonlinear spin Hall effect\cite{Hayami2022}, the gyrotropic Hall effect\cite{Konig2019}, the Magnus Hall effect\cite{Papaj2019}, and the nonlinear Nernst effect\cite{Zeng2019, Liu2025}. 

Similarly, the nonlinear Edelstein effect (NLEE) is an extension of the conventional Edelstein effect, moving beyond linear responses to explore how higher powers of an applied electric field or light can generate spin polarization\cite{Vignale2016, Baek2024, Xu2024, Oike2024, Xu2025}.
The familiar linear Edelstein effect describes how an electric current creates a non-equilibrium spin polarization directly proportional to the current. 
In contrast, the NLEE involves a response that is quadratic or even higher-order on the applied electric field or light intensity. 
A key distinction of the NLEE is its symmetry requirements. While the linear effect typically needs broken spatial inversion symmetry, the NLEE can surprisingly occur in materials that do possess spatial inversion symmetry (centrosymmetric materials). This significantly expands the range of suitable materials to include insulators and semiconductors. The mechanism in these cases might involve electric-field-induced orbital Rashba textures. When considering time-reversal symmetry, the NLEE can even induce magnetization under linearly polarized light in nonmagnetic materials. This phenomenon can be attributed to time-reversal symmetry breaking resulting from energy dissipation through interband transitions\cite{Ye2024, Baek2024}.
The potential implications and applications of the NLEE are significant. It offers a method for electrically generating and controlling spins, potentially leading to novel spintronic devices that overcome the limitations of linear responses. Its ability to create spin or orbital polarization in centrosymmetric materials could enable "field-free" magnetization switching in ferromagnets, which is highly desirable for energy-efficient memory and computing. Optical approaches to NLEE could also provide ultrafast, noncontact, and noninvasive control of magnetism. Furthermore, the relaxed symmetry requirements broaden the material scope for spintronic applications to include common semiconductors and insulators.

The typical perturbative series expansion of the density matrix under electric field interaction yields expressions for the density matrix perturbation at each order \cite{KamalDas}

\begin{equation}
    \rho^{(n+1)}_{I}(t) = -\frac{ie}{\hbar}\int_{-\infty}^{t} dt'e^{ i\mathcal{H}_{0} t' / \hbar }\,\boldsymbol{E}(t') \cdot [\hat{\boldsymbol{r}}, \rho^{(n)}(t')] e^{ -i \mathcal{H}_{0} t' / \hbar }
\end{equation}
where $\rho_I(t)^{(n+1)}$ represents the $(n+1)$-order density matrix in the interaction picture, influenced by the preceding order.
Our attention will be on the second-order density matrix, particularly concerning the intrinsic (impurity-independent) components. This allows us to express the induction of current and spin resulting from the electric field as:
\begin{align}
j_{a} &= -e\sum_{m,n} \rho^{(2)}_{mn} v_{a,nm} = \sigma_{a,bc} E_{b}E_{c} \label{eq:NLHE}
\\
\delta s_{a} &= \sum_{m,n} \rho^{(2)}_{mn} s_{a,nm} = \chi_{a,bc} E_{b}E_{c} \label{eq:NLEE}
\end{align}

As demonstrated in \cite{KamalDas}, the intrinsic conductivity's reliance on the Quantum Geometry Tensor, $\mathcal{Q}$, is significant and can be expressed concisely as follows
\begin{equation}
    \sigma^{\mathrm{int}}_{a,bc} = - \frac{e^{3}}{\hbar} \sum_{\boldsymbol{k}, m \neq n} f_{m} \left[ \partial_{a} \frac{\mathcal{G}_{mn}^{bc}}{\epsilon _{mn}} - 2\left(    \partial_{b} \frac{\mathcal{G}_{mn}^{ac}}{\epsilon _{mn}} + \partial_{c} \frac{\mathcal{G}_{mn}^{ab}}{\epsilon _{mn}} \right)\right]\,,
\end{equation}
In this context, $\epsilon_{mn} = \epsilon_{m} - \epsilon_{n}$ and $f_{m}$ represents the Fermi-Dirac distribution specific to the $m$-th system eigenvalue. The derivative $\partial_{a}$ is defined as $\frac{\partial}{\partial k_{a}}$. The quantum metric $\mathcal{G}_{mn}^{bc}$ is the real part of the quantum geometric tensor $\mathcal{Q}_{mp}^{bc} = \mathcal{R}_{pm}^{b} \mathcal{R}_{mp}^{c}$, where $\mathcal{R}_{mp}^{a} = i\langle u_{m} \vert \partial_{a} u_{p} \rangle$. The compact formulation arises from linking the velocity operator to the Berry connection $\mathcal{R}_{mp}^{a}$. Conversely, as the spin operator lacks a clear connection to the Berry connection, there is no succinct expression for the Edelstein effect using $\mathcal{G}$.

\sectioncustom{Tilted Massive Dirac Model} 
\begin{figure}[h!]
    \centering
    \includegraphics[width=1\linewidth]{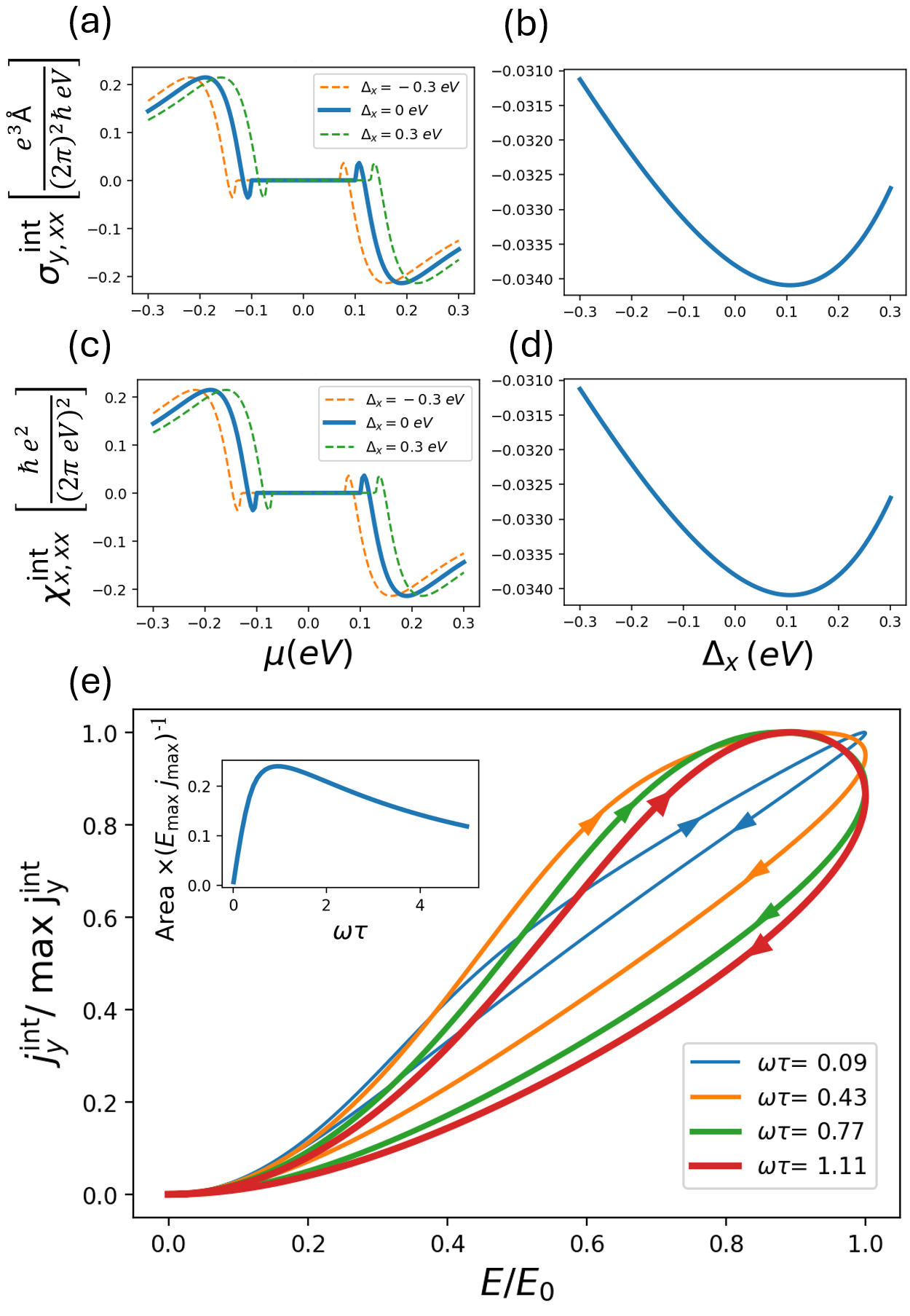}
    \caption{(a), (b), (c) and (d) show the intrinsic conductance and magnetic susceptibilities of the Titled Massive Dirac Model, in function of the chemical potential $\mu$ and $\Delta_x$. In (b) and (d) $\mu = 0.2\; eV$. (e) The current vs electric field cycle is performed using Eq.(\ref{eq:phenomenological}) for various values of $\omega\tau$, and the inset is the plot of the normalized area of the hysteresis loop as a function of $\omega\tau$. The parameters used are $v_F=1 \, eV$\AA, $v_t=0.1 \, eV$\AA, $\Delta_z = 0.2\,eV$, $J = 100\; eV$\AA$^2/\hbar$ and $E_0 = 1\; eV/$\AA.}
    \label{fig:cycleDirac}
\end{figure}
We illustrate the mechanism with the simplest model capable of showing NLHE and NLEE simultaneously.  
We will perform our calculations for a magnetic tilted massive Dirac model, as in \cite{KamalDas}, but generalized by the inclusion of a possible magnetization along the $x$-axis.

\begin{equation}
\mathcal{H} = v_F (k_x \sigma_y - k_y\sigma_x) + \boldsymbol{\Delta}\cdot\boldsymbol{\sigma} + v_t k_y\sigma_0,
\end{equation}
where $v_F$ denotes the Fermi velocity, $v_t$ is the parameter that breaks $\mathcal{PT}$ symmetry and provoke the tilting of the Dirac cone. The $\boldsymbol{\Delta} = (\Delta_x, 0, \Delta_z)$ parameter are possible magnetizations, and $\Delta_z$ opens the gap in the bands. The magnetization on the $x$-axis will play a crucial role due to the induction of spin polarization determined by the Edelstein Effect.

In this simple model, we can perform analytical calculations for the nonlinear response. The calculation of the intrinsic conductivity resembles the calculation performed in \cite{KamalDas}.

\begin{equation}
    \sigma^{\mathrm{int}}_{y,xx} = -\frac{e^{3}\pi v_{t}}{8(2\pi)^{2}\hbar} \frac{1}{4\tilde{\mu}^{2}} \left(  11 - 26 r^{2} + 15 r^{4}\right) 
\end{equation}

where $r = \Delta_{z}/ \tilde{\mu}$ and $\tilde{\mu} = \mu - \Delta_{x}v_{t}/v_{F}$

The connection between angular momentum and linear momentum, given by the SOC coupling in $\mathcal{H}$, allows us to find a similar expression of the NLEE as
\begin{equation}
    \chi^{\mathrm{int}}_{x,xx} = - \frac{\hbar}{2} \frac{e^{2}\pi v_{t}}{8(2\pi)^{2} v_{F}} \frac{1}{4\tilde{\mu}^{2}} \left(  11 - 26 r^{2} + 15 r^{4}\right) 
\end{equation}

The dependence on $\Delta_x$ of these expressions, encapsulated in $\tilde{\mu}$, is the key to the memory effect in the system. We will apply the following scheme in our calculations of the current over a cycle. First, we apply an electric field in the $x$-direction, which will generate a current in the transversal component due to $\sigma_{y, xx}$, but will also generate a magnetization in the $x$-direction. This magnetization will change the Hamiltonian by changing the $\Delta_x$ parameter; this modification can be cast in terms of an effective exchange-correlation potential by $\delta\mathbf{\Delta}=J\delta\mathbf{s}$\cite{Nunez2006, Nunez2006a, Haney2008, Duine2007}. Along a time-dependent sequence, both the NLHE and NLEE will change their values. The actualization of $\Delta_x$ will obey the following phenomenological dissipation equation,
\begin{equation}
    \frac{d}{dt} \Delta_x = - \frac{\Delta_x-(\Delta_x^{(0)} + J\,\delta\mathbf{s})}{\tau}
    \label{eq:phenomenological}
\end{equation}
where $\Delta_x^{(0)}$ is the equilibrium value of the magnetization, which will be set to zero for a paramagnetic material, and we introduced a phenomenological parameter $\tau$ that determines the life expectancy of the excitations that the NLEE generates. We note that, physically, Eq. (\ref{eq:phenomenological}) describes a relaxation mechanism encoded in the many-body response function of the electronic degrees of freedom\cite{Rossi2005, Caldeira1983, Suhl1998}. Its dependence on disorder, phononic, and magnonic excitations will be studied elsewhere. For now, it suffices to regard $\tau$ as a phenomenological parameter and $J$ a conversion constant, from magnetization density to energy, which depends on the interaction of the material and the size of the unit cell. With this, we perform a cycle in electric-field-space ($E_x(t) = E_0 \sin(\omega t)$) and show the presence of memory in our system and the dependence on the phenomenological parameter. For the model to make sense as a viable memory, the frequency must be chosen in the range $\hbar/\tau\lesssim \hbar\omega\ll \mathcal{E}_g$, where $\mathcal{E}_g$ is the electronic gap of the system.

\sectioncustom{Application to CuMnAs}
CuMnAs\cite{Wadley2016} is a fascinating material that's gaining attention for two main reasons: its potential in spintronic applications and its usefulness in studying antiferromagnetic (AFM) Dirac materials\cite{Stamenova2017, Smejkal2017, Saidl2017, Veis2018, Smejkal2018, Tang2016, Linn2023}.
 CuMnAs has a Néel temperature of approximately 480 K. Direct imaging of the Néel vector's reorientation in CuMnAs was achieved through photoemission electron microscopy, with x-ray magnetic linear dichroism supplying the contrast\cite{Grzybowski2017}.

This material exhibits a symmetry known as $\mathcal{PT}$ (combined inversion and time-reversal symmetry)\cite{Smejkal2017, Maca2017, Xu2020}. This symmetry links its two magnetic manganese (Mn) sublattices, which are oriented in opposite directions. Double degeneracy of energy bands is a direct consequence of this $\mathcal{PT}$ symmetry, combined with opposite magnetic moments on inversion partners.
The model we use for CuMnAs is the same as in \cite{KamalDas, Wang2021}. To demonstrate the Hall-memristor idea, we start with a simplified model based on the tetragonal CuMnAs AF, focusing solely on the Mn atoms. Each Mn atom has one orbital, forming stacked, wrinkled, quasi-2D square lattices. We initially disregard the inter-plane coupling due to the larger spacing compared to the distances between first and second nearest neighbors within the planes. In momentum space, the model's Hamiltonian consists of two sublattices with opposite magnetizations, indicating the material's antiferromagnetic nature. Moreover, spin-orbit coupling introduces spin-non-conserving elements. The effective Hamiltonian is represented by a $4 \times 4$ matrix\cite{Smejkal2017}:   

\begin{equation}    
\mathcal{H}(\boldsymbol{k}) = \begin{pmatrix}
\epsilon_{0}(\boldsymbol{k}) + \boldsymbol{h}_{A}(\boldsymbol{k})\cdot \sigma & V_{AB}(\boldsymbol{k}) \\
V_{AB}(\boldsymbol{k}) & \epsilon_{0}(\boldsymbol{k}) + \boldsymbol{h}_{B}(\boldsymbol{k}) \cdot \sigma
\end{pmatrix}
\label{eq:Hcumnas}
\end{equation}
where $\epsilon_{0}(\boldsymbol{k}) = -t(\cos k_{x} + \cos k_{y})$ and $V_{AB}(\boldsymbol{k}) = - 2 \tilde{t} \cos (k_{x} / 2) \cos(k_{y} / 2)$, with $t$ representing the hopping between the same sublattice and $\tilde{t}$ the hopping between different sublattice. The magnetization fields are defined as $\boldsymbol{h}_{A}(\boldsymbol{k}) = -\boldsymbol{h}_{B}(\boldsymbol{k})$ with $\boldsymbol{h}_{A}(\boldsymbol{k}) = (h_{\mathrm{AFM}}^{x} - \alpha _R \sin k_{y} + \alpha_{D}\sin k_{y}, h^{y}_{\mathrm{AFM}} + \alpha_{R} \sin k_{x} + \alpha_{D}\sin k_{x}, h^{z}_{\mathrm{AFM}})$. The presence of spin-orbit coupling (SOC) is explicit, with $\alpha _R$ and $\alpha_{D}$ being Rashba and Dresselhaus SOC, respectively.

\begin{figure}
    \centering
    \includegraphics[width=1\linewidth]{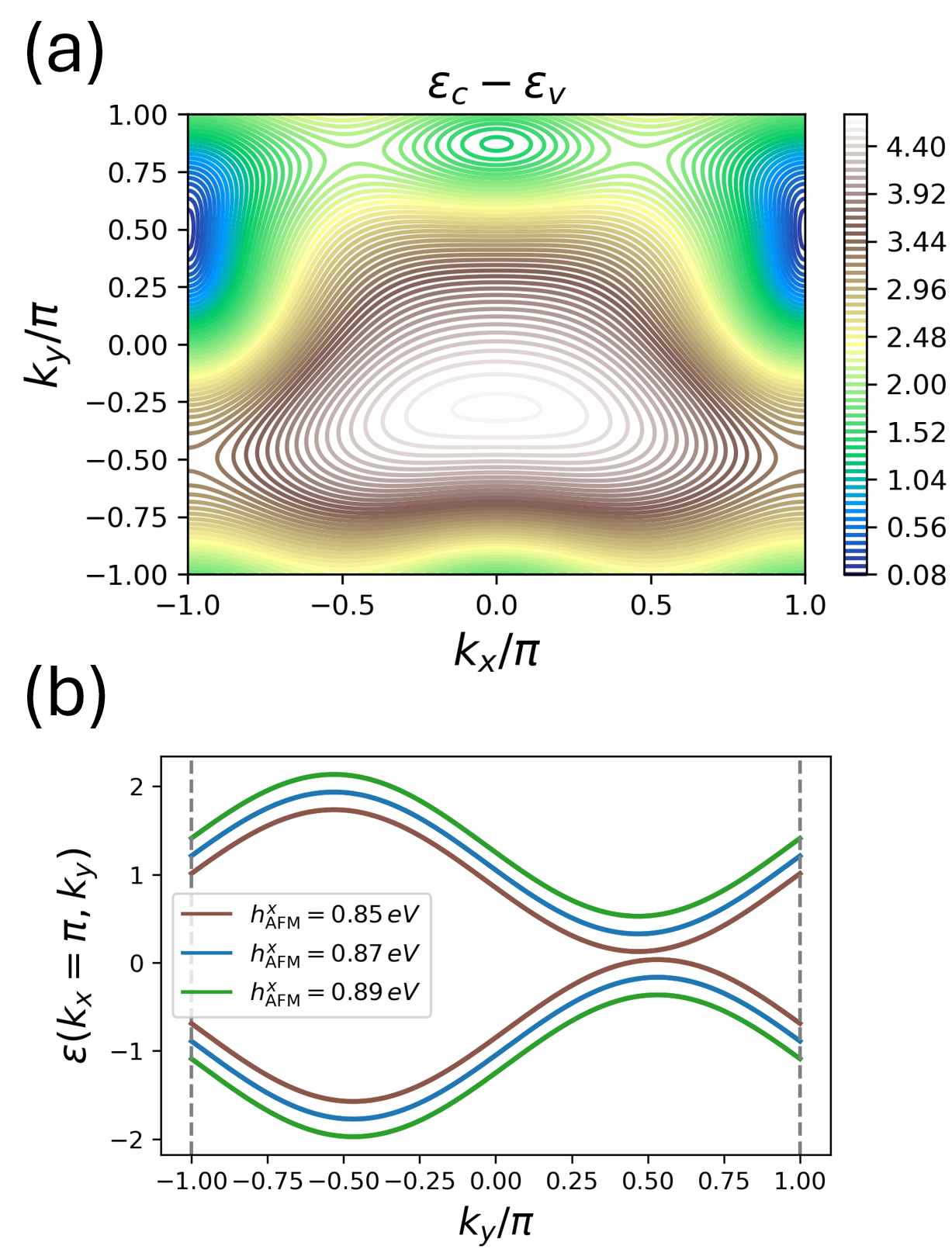}
    \caption{In (a) are the band structure of CuMnAs, it shows that at $\boldsymbol{k} = (1, 0.5)\pi$ are the avoided crossings, whose vicinity  mostly contributes to the NLHE and NLEE. In (b) we can see the $k_y$ dependence of this avoided crossings.}
    \label{fig:energyCuMnAs}
\end{figure}

We perform the same cycle as in Fig.(\ref{fig:cycleDirac}) but now the NLEE distorts the staggered magnetization field, this is $s_{a,nm} = \frac{\hbar}{2} \langle u_n \vert \sigma_z \otimes \sigma_a\vert u_m\rangle$ in Eq.(\ref{eq:NLEE}), where the first space in the Kronecker product is the sub-lattice space and the second is the spin space. We use Eq.(\ref{eq:phenomenological}) with the substitution $\Delta_x \to h_{\rm AFM} ^x$. In the cycle, we will compute the response of this staggered magnetization due to the varying electric field, and this magnetization will change the $\boldsymbol{h}_{\rm AFM}$ in the Hamiltonian in Eq. (\ref{eq:Hcumnas}), affecting the calculation of the Hall conductivity\cite{Wang2021, Volny2020}.

\begin{figure}
    \centering
    \includegraphics[width=1\linewidth]{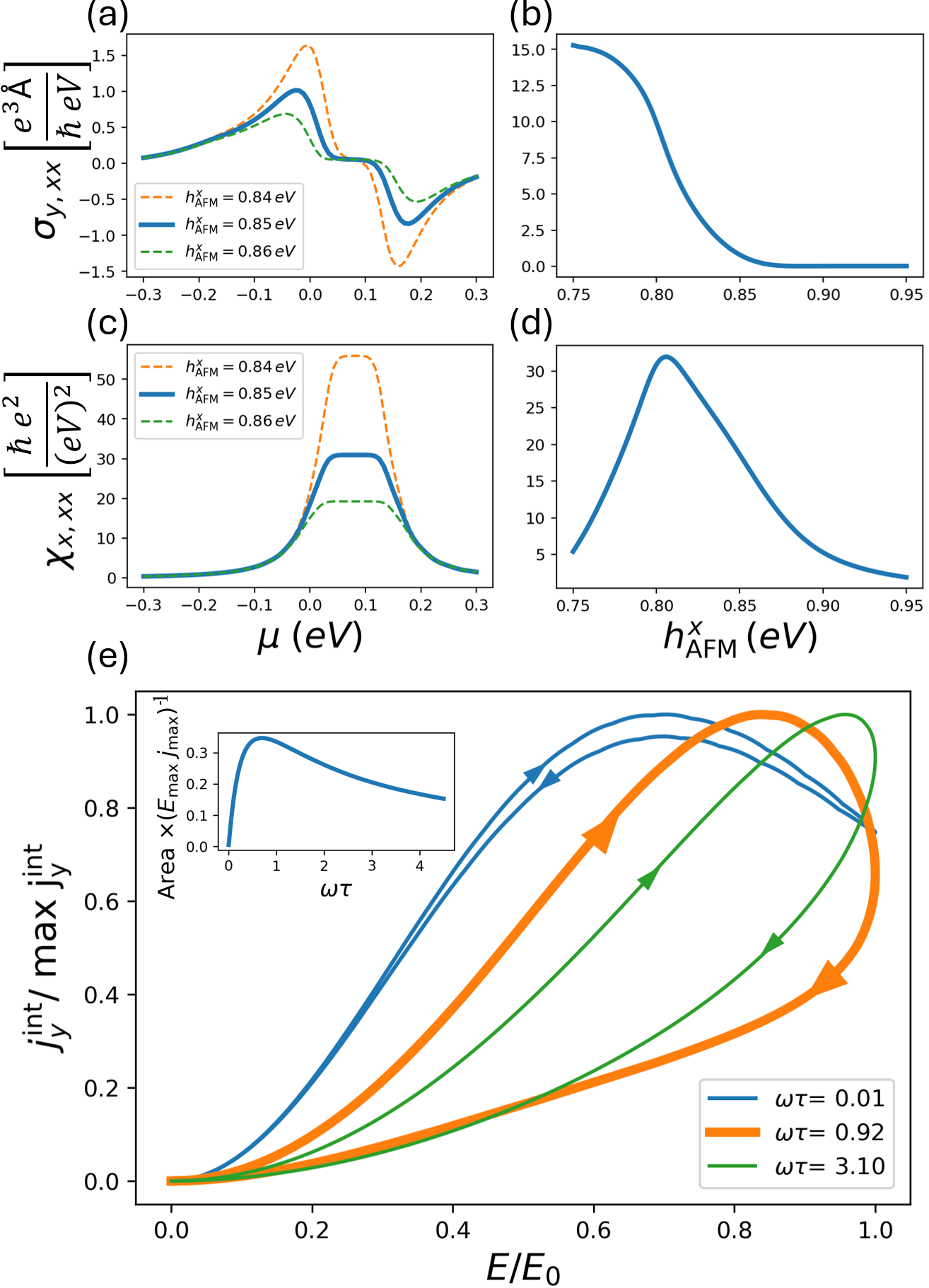}
    \caption{(a),(b),(c) and (d) shows the intrinsic conductance and magnetic susceptibilities for CuMnAs, as a function of the chemical potential $\mu$ and $h^x_{\rm AFM}$. For (b) and (d) $\mu=0\,eV$. (e) The current vs electric field cycle is performed for various values of $\omega\tau$, and the inset is the plot of the normalized area as a function of $\omega\tau$. The value of the parameters are $t = 0.08 \, eV $, $\tilde{t}=1\,eV$, $\alpha_R=0.8 \, eV$, $\alpha_D=0$, $h^y_{\rm AFM}=h^z_{\rm AFM}=0$, $J = 16\; eV$\AA$^2/\hbar$ and $E_0 = 0.01\; eV/$\AA. The full cycle, including negative $E$, is in the Supplementary Material Fig. S1 \cite{SM}.
    }
    \label{fig:cycleCuMnAs}
\end{figure}

\sectioncustom{Discussion} In the context of theoretical memristor analysis, our results offer an extension from the two-terminal memristor setting, customarily the standard kind of device in this context, towards a four-terminal device. The typical two-terminal system provides 4 degrees of freedom, namely current, $I$, voltage, $V$, charge, $Q$, and flux, $\phi$\cite{Chua2019}. The standard memristance is defined as the response relation ${\rm d}\phi/{\rm d}Q=\mathcal{M}(Q)$. In our case, the four-terminal device, in addition to those variables, has associated ones related to the transverse direction. The Hall-memristance, acronym for memory Hall-resistance,  might be defined through ${\rm d}\phi_H/{\rm d}Q={\cal M}_H(\mathbf{n},\mathbf{E})$, where the transverse flux $\phi_H$ is associated with the Hall voltage, $V_H$, through ${\rm d}\phi_H/{\rm d}t=V_H$ and the Hall-memristance depends on the parameter $\mathbf{n}$, the magnetization-like parameter, which in the case of CuMnAs corresponds to the Néel vector, that specifies the magnetic state of the system. We include a dependence on the electric field to model the nonlinear nature of the effect. A standard and straightforward analysis yields the relation: $V_H=\mathcal{M}_H(\mathbf{n},\mathbf{E})\,I$, in agreement with our previous results. Meanwhile, the constitutive relation specifies the rate of change of the Néel vector in response to the electric field through the nonlinear Edelstein effect. We leave the analysis of other cross-responses between the old variables and the new ones (e.g. Hall-memcapacitance or Hall-meminductance) for future explorations.

\begin{figure}
    \centering
    \includegraphics[width=1\linewidth]{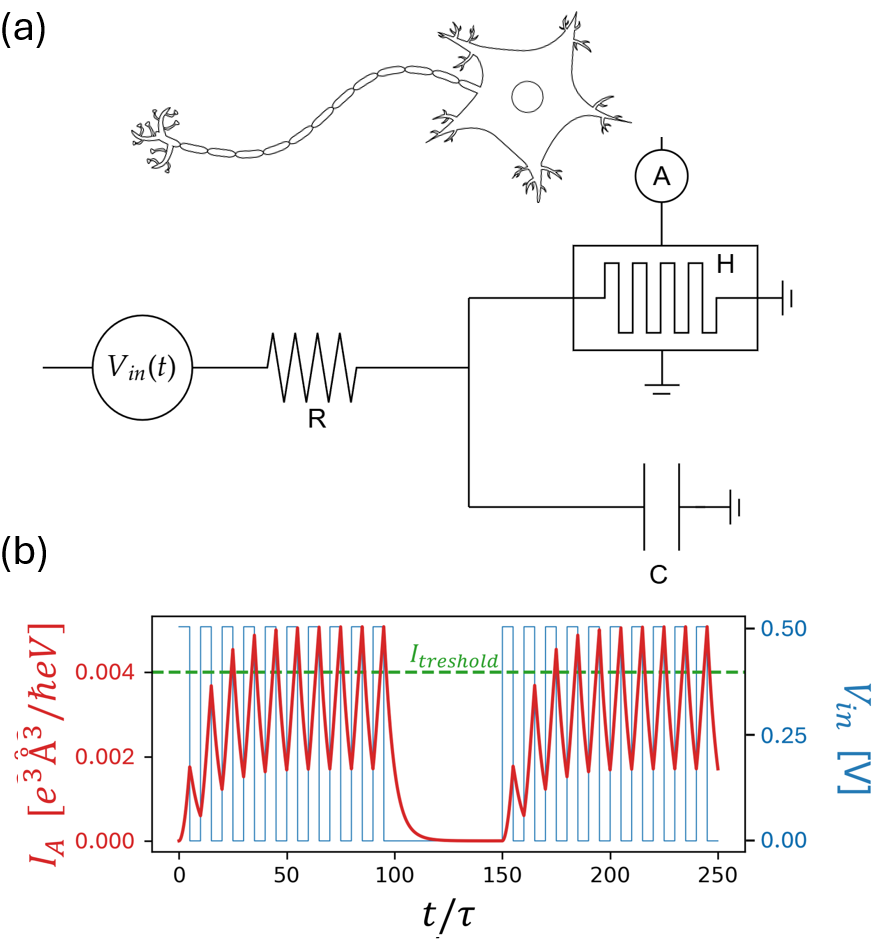}
    \caption{Neuromorphic possibilities for the Hall-Memristor. (a) Proposed circuit for simulating a neuron. (b) output current measure in response to a sequence of pulses. This emulates the connection between neurons, which repeatedly interact, and when a certain number of interactions are made, a pulse is fired. In this memristor, the firing process is possible due to the storage of AF-magnetization that alters the conductivity. More detail in \cite{SM}, Figures S2-S5.}
    \label{fig: neuron}
\end{figure}

In Fig. (\ref{fig: neuron}), we illustrate the possibilities of a Hall-memristor in the implementation of artificial neurons \cite{park2022}. We have denoted the four-terminal Hall memristor device with the same symbol as a memristor with an 'H' on the body.

\sectioncustom{Conclusions}  In this study, we introduced a spin-memristor using antiferromagnetic materials that display Hall-memresistance. The effect is based on the coexistence of the nonlinear Hall and Edelstein effects. The nonlinear Hall effect plays the role of a reader of the memory stored in the Néel vector properties. The Edelstein effect serves as the recorder of the information. We performed a symmetry-based analysis to demonstrate the feasibility of the effect. Focusing on CuMnAs, known for its controllable nonlinear Hall effect, we aimed for a practical realization of these concepts. 
The advancement of spin-memristors is vital for overcoming the limitations of silicon electronics, enabling intelligent and energy-efficient computing. Our proposed antiferromagnetic Hall-memristor offers key benefits, including ultra-fast dynamics, high integration density without stray magnetic fields, and resilience to external magnetic fields. The interplay between nonlinear Hall and Edelstein effects in our device facilitates efficient electrical control and detection of spin states, a significant challenge in antiferromagnetic spintronics.
We emphasize that our findings extend the typical two-terminal spin-memristors to include a four-terminal device.

\sectioncustom{Acknowledgements} Funding is acknowledged from Fondecyt Regular 1230515 and Cedenna CIA250002. 

\bibliography{memristor}
\end{document}